\documentclass[prl, aps, twocolumn, superscriptaddress]{revtex4}

\usepackage{graphicx, epsfig}
\usepackage{amssymb,amsmath}

\def\lba{\left(}    \def\rba{\right)}
\def\lbc{\left[}    \def\rbc{\right]}

\newcommand{\ket}[1]{\left|{#1}\right.\rangle}

\newcommand{\vk}{{\bf k}}  
\renewcommand{\vr}{{\bf r}}

\begin{document}

\title{Trapped fermionic clouds distorted from the trap shape due  
to many-body effects}

\author{Masudul Haque}

\affiliation{Max Planck Institute for Physics of Complex Systems,
  Dresden, Germany}

\affiliation{Institute for Theoretical Physics, Utrecht University,
Leuvenlaan 4, 3584 CE Utrecht,
The Netherlands}

\author{H.~T.~C.~Stoof}

\affiliation{Institute for Theoretical Physics, Utrecht University,
Leuvenlaan 4, 3584 CE Utrecht,
The Netherlands}

\date{\today}

%
%

\begin{abstract}

We present a general approach for describing trapped Fermi gases, when the
cloud shape is distorted with respect to the trap shape. Our approach provides
a consistent way to explore physics beyond the local density approximation, if
this is necessary due to the distortion.  We illustrate this by analyzing in
detail experimentally observed distortions in a trapped imbalanced Fermi
mixture.  In particular, we demonstrate in that case
dramatic deviations from ellipsoidal cloud shapes arising from the competition
between surface and bulk energies.

\end{abstract}


\maketitle

\emph{Introduction.} ---
The cooling of trapped fermionic atoms to degeneracy has opened up a
fascinating new arena for studying many-fermion physics
\cite{fermionic-cooling-expts}.
In particular, the opportunity to access the BEC-BCS crossover using
Feshbach resonances has led to a large effort in the experimental
study of fermionic pairing using trapped atomic gases
\cite{BCS-BEC-expts}.
Among other topics, pairing in polarized Fermi gases is now being intensely
explored, experimentally in Refs.~\cite{Ketterle_1st-paper, Hulet_1st-paper,
Ketterle_2nd-paper, Hulet_2nd-paper} and theoretically in, e.g.,
Refs.~\cite{SheehyRadzihovsky_PRL06, Caldas_PRA04, Chevy_unequal-popln-thy,
Haque-Stoof-LDA, Mueller-LDA, Mueller_surf-tens_PRL06}. 
Apart from possible new insights for solid-state and nuclear physics, the
field of trapped fermionic atoms comes with its own set of new concerns, such
as the role of the trapping potential.
In this Letter, we therefore provide results on the relationship between the
trap shape and the shape of the gas cloud.

For a large class of trapped many-body systems, the local density
approximation (LDA) gives an accurate description of the spatial
distributions of atomic and energy densities.
However, motivated by recent experiments \cite{Hulet_1st-paper,
Hulet_2nd-paper}, we propose that a number of mechanisms may lead to
the intriguing situation that the combination of many-body physics and
the external trapping potential causes the shape of the gas cloud to
be distorted compared to the shape determined by the trap alone.  In
such cases, the simple LDA procedure in terms of a local chemical
potential is no longer appropriate, and we need to consider the
angular structure
of the local Fermi surface. We have therefore formulated an extension
of the LDA that allows for a distinction between the different
directions.
We give below some examples where one observes or expects such spatial
distortions.  We note however that there might be many other situations,
beyond what we anticipate here, where such distortions take place.  Our
formalism is thus of very general applicability and interest.

As a first scenario for spatial distortions, consider a many-body
system that is phase separated in an elongated trap due to the
presence of a first-order transition.  The interface between the two
phases then carries a surface tension.  To minimize the surface
energy, the interface shape will differ from the trap shape.  In fact,
such a system has recently been experimentally realized using a
trapped polarized Fermi mixture of resonantly-interacting atoms
\cite{Hulet_1st-paper, Hulet_2nd-paper}. At sufficiently low
temperatures this mixture phase separates into an unpolarized
superfluid core and an outer shell of normal polarized fermions
\cite{Hulet_1st-paper, SheehyRadzihovsky_PRL06, Caldas_PRA04,
Chevy_unequal-popln-thy, Mueller-LDA, Haque-Stoof-LDA}. When the trap
is elongated enough and the particle number is small enough, the
surface tension causes significant deviation of the core aspect ratio
from the trap aspect ratio \cite{Mueller_surf-tens_PRL06,
Hulet_2nd-paper}. In the following we will analyze this situation in
detail.

Second, Pomeranchuk instabilities are shape deformation instabilities
of a Fermi surface \cite{Pomeranchuk_JETP58}, which have attracted
intense interest in the strongly correlated electrons community
\cite{pomeranchuk_thy-refs,QuintanillaSchofield_cm06}.
%
%
A $d$-wave
Pomeranchuk transition leads to a ``nematic'' Fermi liquid with an
elliptical Fermi surface. Other than their possible realization in
solids, it is hoped that such a nematic liquid may be realized in
atomic dipolar Fermi gases since a long-range interaction can cause a
Pomeranchuk instability if the short-range part is suppressed
\cite{QuintanillaSchofield_cm06}. A Fermi surface deformation that is
constant in space causes no distortion in coordinate space. However,
in an
external trap, the Fermi surface deformation would vary from point to
point, leading to a distortion of the gas cloud in coordinate space as
well. To calculate particle and energy densities in such a case, would
again require an anisotropic generalization of the LDA, of the kind we
provide here.

Finally, one possible mechanism for pairing of polarized fermions is
via opposite deformation of the two Fermi surfaces
\cite{Sedakrian_Deformed-FS-pairing}. A prolate (oblate) shape for the
majority (minority) Fermi surface leads to an equatorial region in
momentum space along which opposite-momenta fermions can pair. With
the extended parameter space expected to be accessible with trapped
atoms in the future, it is
likely that this pairing mechanism is energetically favorable in some
parameter regime \cite{Sedakrian_Deformed-FS-pairing}. In the case of
such pairing in a trap, we expect both Fermi surfaces to have
ellipticities (of opposite sign) that vary with position.  As a
result, the majority and minority clouds are expected to be distorted
in opposite directions relative to the trap.  Again, a treatment of
the type presented in this Letter is needed to describe this
situation.

\emph{Anisotropic generalization of local density approximation}--- 
The LDA encodes the inhomogeneity of the gas completely in terms of a
local chemical potential at every point in space, $\mu(\vr) = \mu -
V_{\rm tr}(\vr)$. Here $\mu$ is the true chemical potential of the gas
and $V_{\rm tr}(\vr)$ is the trapping potential. The shape of the gas
cloud is determined by an equipotential surface of the trap.
For harmonic traps, this surface is an ellipsoid.

\begin{figure}
\centering
\includegraphics*[width=0.95\columnwidth]{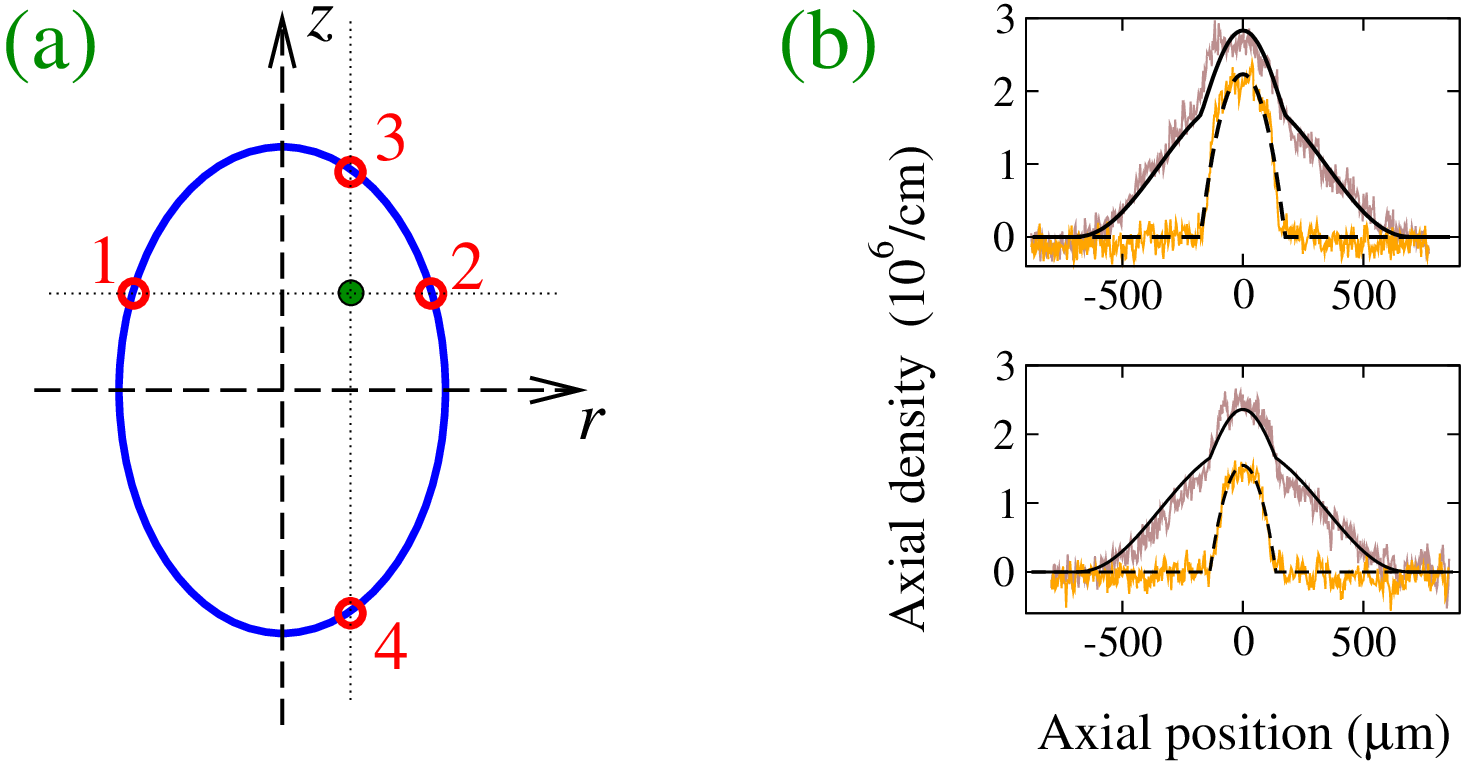}
\caption{  \label{fig_prescription-n-axial-fits}
(Color online.)  (a) Calculation of the local anisotropic Fermi
  surface at position $\vr$ (full dot). Points 1 and 2 with
  coordinates $({\pm}R(z),z)$ are the classical turning points for the
  motion in the $r$-direction. Points 3 and 4, i.e., $(r,{\pm}Z(r))$,
  are the classical turning points for motion in the $z$
  direction. (b) Fits to the axial density profiles for both species
  using our approach.  The top and bottom panels show two experimental
  shots, at polarizations $P\simeq{0.5}$ and $P\simeq{0.65}$,
  respectively.
}
\end{figure}

The key to treating distorted systems is to generalize the
definition of the local chemical potential $\mu(\vr) = \mu -
V_{\rm tr}(\vr)$ in a way that discriminates between directions.
For definiteness, we will consider axially symmetric traps,
$V_{\rm tr}(\vr) = V_{\rm tr}(r,z)$, where $r$ and $z$ are the
radial and axial length variables of a cylindrical coordinate
system, respectively. 
%
%
For each position in the trap, we need to determine the momentum-space
location of the deformed Fermi surface, in both the radial and axial
directions.  We denote by $R(z)$ and $Z(r)$ the edge of a reference
ideal Fermi cloud, that would be obtained by hypothetically turning
off the interactions but retaining the distortion.  Our prescriptions
for the location of the Fermi surface at position $\vr$ in the radial
and axial directions are
\begin{subequations} \label{eq_improved-LDA}
\begin{align}
\mu_{\rm R}(\vr) ~&=~ \mu_{\rm R}(r,z) ~=~ V_{\rm tr}(R(z),z) -
V_{\rm tr}(r,z)~,
\\
\mu_{\rm Z}(\vr) ~&=~ \mu_{\rm Z}(r,z) ~=~  V_{\rm tr}(r,Z(r)) -
V_{\rm tr}(r,z)~.
\end{align}
\end{subequations}
We have written the Fermi surface locations in energy units,
i.e., $\mu_{\rm R}$ and $\mu_{\rm Z}$ are the free-particle energies
corresponding to the Fermi momenta in the radial and axial directions.
They are, of course, not components of the chemical
potential. Eqs.~\eqref{eq_improved-LDA} are obtained by semiclassical
considerations similar to the WKB approximation, which is one way to
derive the LDA.  Here, we have applied the semiclassical approximation
separately to the $r$ and $z$ directions
(Fig.~\ref{fig_prescription-n-axial-fits}a) as appropriate for an
axially symmetric trap, where the single-particle Schr\"odinger
equation separates in these two directions.

After obtaining the locations of the local Fermi surface in the radial and
axial directions, it is for many purposes sufficient to use an ellipsoidal
Fermi surface with $\mu_{\rm R}$ and $\mu_{\rm Z}$ as the principal axes. At
the center of the trap, the same semiclassical argument can be carried out in
any direction and the local Fermi surface has exactly the same shape as the
edge of the ideal gas reference cloud. The cloud shape is generally near
ellipsoidal but may deviate from this shape either due to anharmonicity of the
trap or due to many-body effects, as we will see. Although the shape of the
Fermi surface away from the trap center is not fixed by similar arguments, it
is physically reasonable to use the same shape as in the trap center.

The local non-spherical Fermi surfaces obtained using this
prescription can now be used to calculate densities. For example, for
an elliptically distorted normal gas in a harmonic trap, the density
is given in terms of the volume of the local Fermi surface by $n(\vr)=
\lbc(2m)^{3/2}/6\pi^2\rbc \mu_{\rm R}(\vr)\sqrt{\mu_{\rm Z}(\vr)}$.
This density vanishes at the surface defined by $R(z)$ or $Z(r)$.  For
a superfluid, on the other hand, it may be appropriate to use the BCS
equations generalized to anisotropic Fermi surfaces, i.e., the
momentum integrand in the density profile
\[
n(\vr) = \int_0^{\infty} \frac{k^2dk}{2\pi^2} \int_{-1}^{1} dx
\lba 1-\frac{\epsilon_\vk-\epsilon_{\rm
F}(x)}{\sqrt{[\epsilon_\vk-\epsilon_{\rm F}(x)]^2+|\Delta|^2}}\rba
\]
now has angular dependence. Here $x=\cos\theta$, $\epsilon_\vk =
\hbar^2\vk^2/2m$ with $m$ the atomic mass, $\Delta$ is the gap, and
$\epsilon_{\rm F}(x)$ is the deformed Fermi surface with principal
axes $\mu_{\rm R}$ and $\mu_{\rm Z}$.  Similar modifications can be
made for the BCS energy equation and the BCS gap equation.

\emph{Application to polarized resonantly-interacting Fermi gases} ---
We now demonstrate our procedure by elaborating on a particular case,
i.e., we analyze the experiment reported in
Ref.~\cite{Hulet_2nd-paper} using
the formalism presented above.
Figure 2 of Ref.~\cite{Hulet_2nd-paper} shows that the majority aspect ratio
does not change much with polarization. We therefore describe the majority
species in state $\ket{\uparrow}$ with the usual LDA. On the other hand, the
boundary of the minority atom cloud in state $\ket{\downarrow}$ is distorted
due to surface tension. We thus treat the minority according to the approach
described above.
At each position within the minority cloud, we calculate the two parameters
for the Fermi surface of the minority species, using as an input the boundary
of the minority cloud, which coincides with the superfluid core if any
intermediate partially-polarized normal shell can be neglected.

For a choice of $\mu_{\uparrow}$, $R_{\downarrow} \equiv
R_{\downarrow}(0)$, and $Z_{\downarrow} \equiv Z_{\downarrow}(0)$, the
particle densities $n_{\uparrow}(\vr)=n_{\downarrow}(\vr)$ and the
energy density $e(\vr)$ at any point inside the core can be calculated
from the local Fermi surfaces. 
The majority Fermi surface is given by $\mu_{\uparrow}(\vr) =
\mu_{\uparrow}-V_{\rm tr}(\vr)$.
The minority Fermi surface is deformed, with
$\mu_{R\downarrow}(\vr)$ and $\mu_{Z\downarrow}(\vr)$ calculated using
Eqs.~\eqref{eq_improved-LDA}.  
Since at resonance the gap is larger than the maximum difference
between the Fermi surfaces, the minority momentum distribution is not
anisotropic even though $\mu_{R\downarrow}\neq\mu_{Z\downarrow}$,
unlike the case of Refs.~\cite{Sedakrian_Deformed-FS-pairing}.
For the fully polarized normal gas outside the core,
$n_{\uparrow}(\vr)$ and $e(\vr)$ are calculated by the usual LDA.

To calculate the densities and superfluid gap in the superfluid core,
we could use generalized BCS equations as described above.  
%
%
However, with one additional approximation, we can
also make use of Monte Carlo results \cite{CarlsonReddy_PRL05} for the
resonant situation near which the experiments of
Refs.~\cite{Hulet_1st-paper, Hulet_2nd-paper} are performed. The
additional approximation is to use at every position an effective
chemical potential $\mu_{\downarrow}(\vr) =
[\mu_{R\downarrow}(\vr)]^{2/3}[\mu_{Z\downarrow}(\vr)]^{1/3}$ for the
minority species as well. The superfluid properties are then given by
the universal unitarity results \cite{CarlsonReddy_PRL05} in terms of
the average chemical potential $\mu(\vr) = [\mu_{\downarrow}(\vr) +
\mu_{\uparrow}(\vr)]/2$.

In Fig.~\ref{fig_prescription-n-axial-fits}b, we show the resulting
axial density profiles, and compare with experimental data for two
different polarizations $P=(N_{\uparrow}-N_{\downarrow})/
(N_{\uparrow}+N_{\downarrow})$.  The calculation requires three
parameters ($\mu_{\uparrow}$, $R_{\downarrow}$, $Z_{\downarrow}$) as
input.  Unlike the calculation in Ref.~\cite{Haque-Stoof-LDA}, the
required input quantities are not immediately fixed by the trap
parameters and particle numbers ($N_{\uparrow}$,$N_{\downarrow}$)
alone. An additional energy calculation is required for the aspect
ratio of the core, as explained later. The experimental majority axial
density profiles have pronounced peaks at the center. This is
reproduced well by the calculations shown in
Fig.~\ref{fig_prescription-n-axial-fits} using Monte Carlo unitarity
parameters, but not if we use the BCS equations to describe the
core. Note that these central peaks are not prominent in the first
experiment by Partridge {\it et al}.~\cite{Hulet_1st-paper}, nor in
the experiment by Shin {\it et al}.\ \cite{Ketterle_2nd-paper},
suggesting that thermal fluctuations play a stronger role in those
cases.  In all calculations reported in this Letter, anharmonicities
of the experimental trap shape are included unless explicitly stated
otherwise.

\emph{Energy minimization} --- At unitarity, the surface energy
density associated with the superfluid-normal interface is expected to
be given by the chemical potentials through universal
constants. Defining $\mu=(\mu_{\uparrow}+\mu_{\downarrow})/2$ and $h =
(\mu_{\uparrow}-\mu_{\downarrow})/2$, we first note that a quantity
with the dimension of a surface energy density constructed out of
$\mu$ alone must be proportional to $(m/\hbar^2)\mu^2$. The dependence
on $h$ may be included as a function of the dimensionless ratio
$h/\mu$.  The surface tension thus has the form $\sigma = \eta_s
(m/\hbar^2) \mu^2 f_{\rm s}(h/\mu)$~. Here $\eta_s$ is a dimensionless
number expected to be of order $1$ and $f_{\rm s}$ is a dimensionless
function.  Noting in addition that the position of the interface is
fixed by the first-order transition determined by a universal value of
$h/\mu$ \cite{CarlsonReddy_PRL05}, we infer that $f_{\rm s}(h/\mu)$ is
a constant along the surface we are interested in. We thus absorb it
into the universal constant $\eta_s$.

\begin{figure}
\centering
 \includegraphics*[width=0.95\columnwidth]{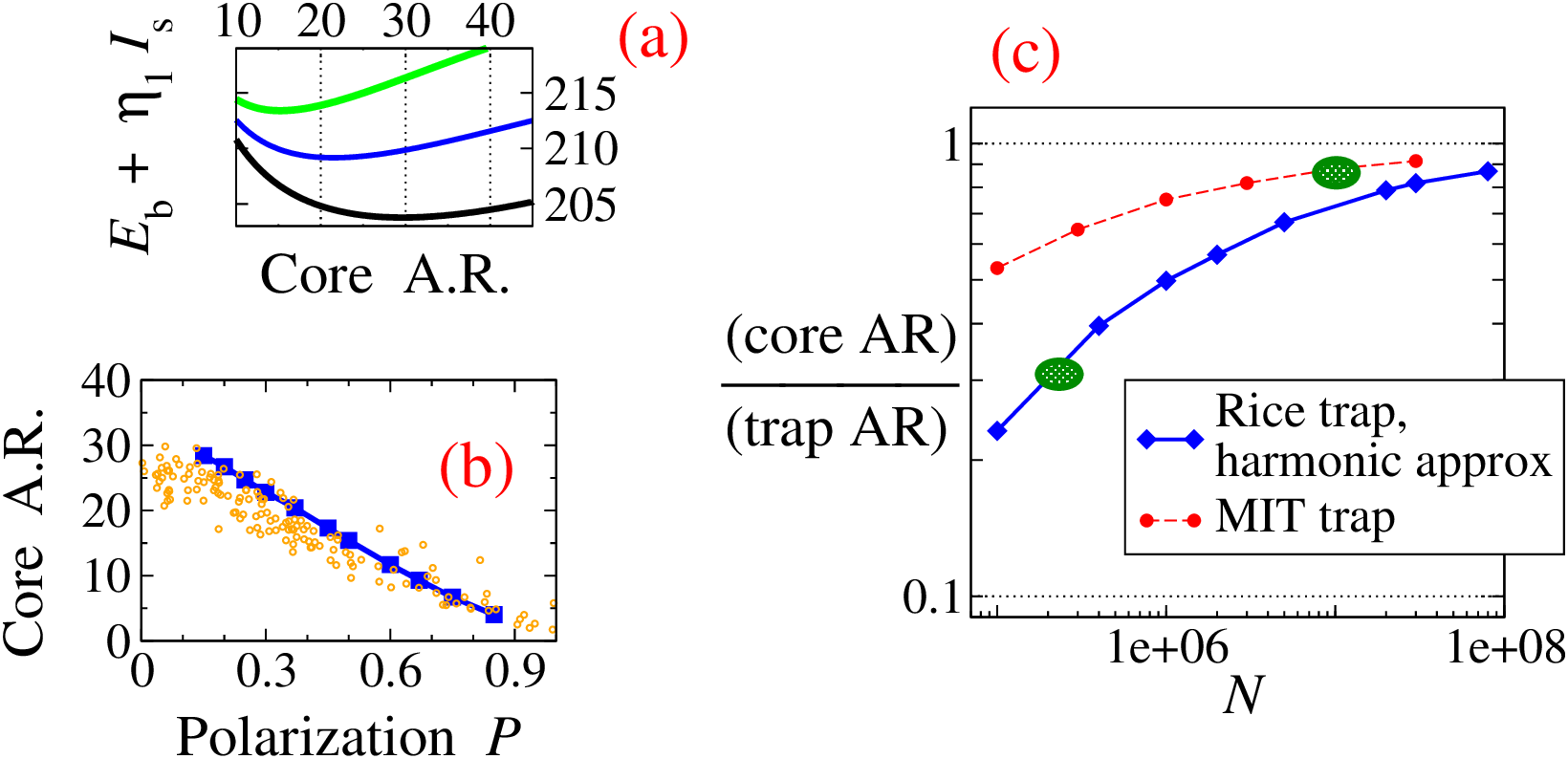}
\caption{  \label{fig_AspectRatios}
(Color online.)
(a) Total energy curves for fixed
  $(N_{\uparrow},N_{\downarrow})=(194,92.7)\times10^3$, shown for
  $\eta_s$ values of 0.30, 0.59 and 0.90 from bottom to top.
(b) Core aspect ratio (A.R.) as a function of polarization for
  $N_{\uparrow}/10^3 = 200-80P$ and calculated
  with $\eta_s = 0.6$.
(c) Core A.R. versus particle number, for fixed polarization $P=0.49$.
  The regions relevant to experiments \cite{Ketterle_1st-paper, Hulet_1st-paper,
  Hulet_2nd-paper, Ketterle_2nd-paper} are indicated by shaded
  ellipses.
}
\end{figure}

We can use the observed core aspect ratios
$Z_{\downarrow}/R_{\downarrow}$ extracted from the axial density data
to determine $\eta_s$, because the correct value of $\eta_s$ must
produce a total energy minimum at the correct aspect ratio. Denoting
by $n_{\rm N}(\mu)$ and $e_{\rm N}(n)$ the particle and energy-density
functions of a homogeneous two-component ideal Fermi gas, the bulk
energy density in the superfluid core is given by $e(\vr)=
(1+\beta)e_{\rm N}\lba n(\vr) \rba + V_{\rm tr}(\vr) n(\vr) $, where
$n(\vr) = n_{\rm N}\lba\mu(\vr)/(1+\beta)\rba$ is the density profile
in the core, $\beta \simeq -0.585$ is a universal constant
\cite{CarlsonReddy_PRL05}, and $\mu(\vr)$ is calculated as described
previously. The energy density in the outer shell is given simply by
the single-component ideal Fermi gas expression.  Integrating over the
core and the outer shell gives the total bulk energy $E_{\rm b}$.

The total energy is $E_{\rm b} + \eta_s I_{\rm s}$, where $I_{\rm
s}$ is the surface integral over $(m/\hbar^2)\mu^2(\vr)$. In
Fig.~\ref{fig_AspectRatios}a, this quantity is shown for several
values of $\eta_{\rm s}$, for fixed populations.
The axial
data in an experimental shot for these populations is fit well
with $Z_{\downarrow}/R_{\downarrow} \simeq 21.5$ and a minimum is
located at this core aspect ratio for $\eta_{\rm s} \simeq 0.59$.
Carrying out this procedure with several experimental axial
density profiles, we obtain the estimate $\eta_{\rm s} =
0.60\pm0.15$. The error bar leads to uncertainties in the
calculated  aspect ratios that is about 2\% for the smallest
deformations and about 10\% for the largest deformations.

\emph{Variation of core deformation} --- Having determined the
universal constant $\eta_{\rm s}$, we can compare our theory to
the experimental core aspect ratio versus polarization curve in
Figure 2 of Ref.~\cite{Hulet_2nd-paper}. The aspect ratio actually
depends on the total particle number in addition to the
polarization. In the experiment, there are shot-to-shot variations
of $N_{\uparrow}$ and a systematic bias towards larger
$N_{\uparrow}$ at smaller $P$. We model this bias as $N_{\uparrow}
= (200-80P)\times10^3$.  Most of the shots are scattered
around this line. Our results are shown in
Fig.~\ref{fig_AspectRatios}b, together with the experimental data.

In Fig.~\ref{fig_AspectRatios}c, we display the core deformation as a function
of total particle number. Surface effects are less important in the
thermodynamic limit $N\rightarrow\infty$ and the difference between core and
trap aspect ratios vanishes as $N^{-1/3}$ in this limit
\cite{Mueller_surf-tens_PRL06}. The lower curve is for a harmonic trap
matching the trap of Ref.~\cite{Hulet_2nd-paper} at the center. The upper
curve presents our expected result for the experiments of Shin {\it et al}.\
\cite{Ketterle_2nd-paper}. The predicted deformation of about 10\% appears to
be sufficiently large to have been observable in these experiments, which is
consistent with the interpretation of Gubbels {\it et al}.\ \cite{Koos}
that no phase separation has occurred in this case.

\begin{figure}
\centering
\includegraphics*[width=0.95\columnwidth]{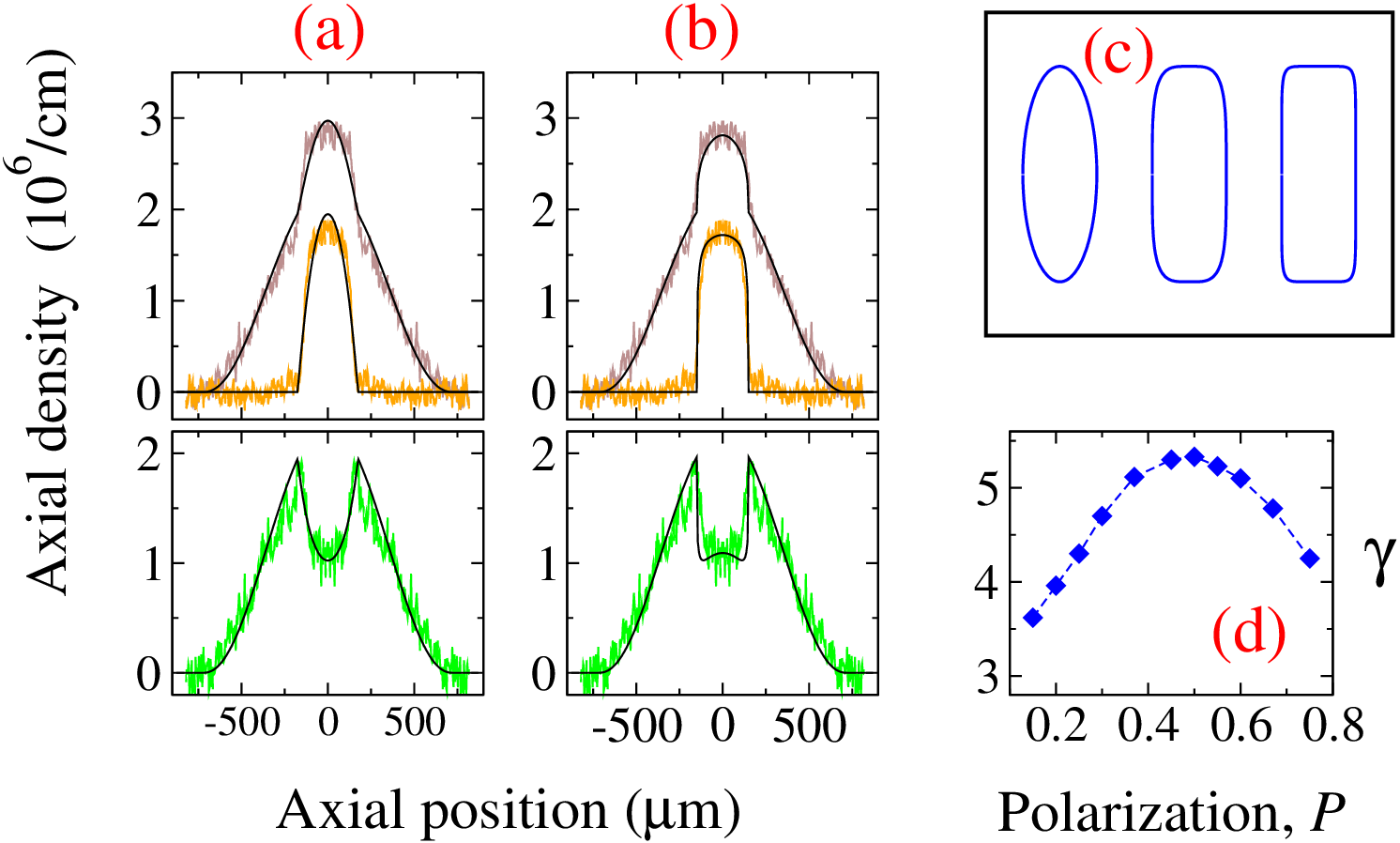}
\caption{  \label{fig_shape-parameter}
(Color online.)  Experimental axial density curves fitted with core shapes (a)
  $\gamma=2$ and (b) $\gamma=8$.  Top panels show species axial densities,
  bottom panels show difference. The value $\gamma=2$ misses the "flat"
  features, while $\gamma=8$ over-emphasizes them.
(c) Surface shapes for $\gamma =$ 2, 4, and 9, respectively.
(d) Calculated value of $\gamma$ for various polarizations, along
the line $N_{\uparrow}/10^3 = 200-80P$.
}
\end{figure}

\emph{Deviation from ellipsoidal shape} --- Examination of the
high-polarization data of Figure 1 of Ref.~\cite{Hulet_2nd-paper} reveals a
peculiar feature. The minority absorption images, other than having a lower
aspect ratio than the trap, also have shapes that are not quite ellipsoidal
and tend to being somewhat cylindrical. The axial density profiles reveal the
same geometric effect in terms of flatter peaks than would be expected for an
ellipsoidal shape.
%
%
Our formalism is well-suited for studying this effect,
since the core boundary simply translates into distortions of local
Fermi surfaces that can be used to calculate consistent density and
energy profiles. We parameterize non-elliptical shapes by describing
the core surface by the equation $(r/R)^{\gamma} + (z/Z)^{\gamma} =
1$. Using a $\gamma$ equal to 2 gives an ellipsoid and infinite
$\gamma$ gives a cylinder.  
Comparing axial density profiles obtained with various values of
$\gamma$, as in Figs.~\ref{fig_shape-parameter}a and
\ref{fig_shape-parameter}b, shows that the experimental features,
e.g.\ the flat tops, are best reproduced with $\gamma$ between 3 and
6, with the best $\gamma$ tending to increase with increasing
polarization, up to about $P\simeq 0.6$.  However, a precise
determination of $\gamma$ is difficult from these fits alone.

Our formalism also allows us to extract $\gamma$ from an energy calculation.
Minimizing the total (bulk plus surface) energy, under the constraint of
constant particle numbers, produces the observed deviation from ellipsoidal
shape, with $\gamma>2$ and $\gamma$ increasing with $P$ for moderate
polarizations (Fig.~\ref{fig_shape-parameter}d).  The energy calculation also
predicts a decrease of $\gamma$ at larger $P$.
This prediction is difficult to verify from the currently available axial
density data because of the large fluctuations of ($N_{\uparrow}$,
$N_{\downarrow}$) from the line $N_{\uparrow}/10^3 = 200-80P$.



\emph{Conclusions} --- To summarize, we have presented an
anisotropic extension of the local density approximation in order
to deal with situations where the aspect ratio of trapped
fermionic gases differs from the trap aspect ratio due to
many-body effects. Our procedure allows for a consistent
calculation of the density and energy profiles. We have
demonstrated this via a detailed analysis of experiments with
phase-separated polarized fermions.

In particular, we have uncovered and explained a striking new effect
concerning the shape of the superfluid core, which takes an unexpected
non-ellipsoidal form. Our treatment dramatically highlights the
utility of our formulation: we know of no other way of calculating the
density distributions for an arbitrary cloud shape.

 
\emph{Acknowledgments} --- We benefited greatly from discussions and
collaboration with the experimental group at Rice University, in particular
Wenhui Li and Randy Hulet.  
%

\end{document}